%% file: main.tex
\documentclass[review,11pt]{elsarticle}
\usepackage{graphicx} 

\usepackage{todonotes}
\usepackage{markdown}
\usepackage{geometry}
\usepackage{array}
\usepackage[numbers]{natbib} %
\begin{document}

\title{De-centering the (Traditional) User: \\ Multistakeholder Evaluation of Recommender Systems}

\author[1]{Robin Burke}
\ead{robin.burke@colorado.edu}
\affiliation[1]{%
  organization={Department of Information Science, University of Colorado, Boulder}, city={Boulder, Colorado}, country={USA}, postcode={80309}
}

\author[2]{Gediminas Adomavicius}
\ead{gedas@umn.edu}
\affiliation[2]{%
  organization={Department of Information and Decision Sciences, University of Minnesota}, city={Minneapolis, Minnesota}, country={USA}
}

\author[3]{Toine Bogers}
\ead{tobo@itu.dk}
\affiliation[3]{%
  organization={IT University of Copenhagen}
  , city={Copenhagen}
  , country={Denmark}
}

\author[4]{Tommaso Di Noia}
\ead{tommaso.dinoia@poliba.it}
\affiliation[4]{%
  organization={Polytechnic University of Bari}
  , city={Bari}
  , country={Italy}
}

\author[5]{Dominik Kowald}
\ead{dkowald@know-center.at}
\affiliation[5]{%
  organization={Know Center Research GmbH \& Graz University of Technology}
  , city={Graz}
  , country={Austria}
}

\author[6]{Julia Neidhardt}
\ead{julia.neidhardt@tuwien.ac.at}
\affiliation[6]{%
  organization={CD Lab for Recommender Systems, TU Wien}
  , city={Vienna}
  , country={Austria}
}

\author[7]{Özlem Özgöbek}
\ead{ozlem.ozgobek@ntnu.no}
\affiliation[7]{%
  organization={Norwegian University of Science and Technology}
  , city={Trondheim}
  , country={Norway}
}

\author[8]{Maria Soledad Pera}
\ead{m.s.pera@tudelft.nl}
\affiliation[8]{%
  organization={TU Delft}
  , city={Delft}
  , country={Netherlands}
}

\author[9]{Nava Tintarev}
\ead{n.tintarev@maastrichtuniversity.nl}
\affiliation[9]{%
  organization={Maastricht University}
  , city={Maastricht}
  , country={Netherlands}
}

\author[10]{Jürgen Ziegler}
\ead{juergen.ziegler@uni-due.de}
\affiliation[10]{%
  organization={University of Duisburg-Essen}
  , city={Duisburg}
  , country={Germany}
}

\begin{abstract}
Multistakeholder recommender systems are those that account for the impacts and preferences of multiple groups of individuals, not just the end users receiving recommendations. Due to their complexity, these systems cannot be evaluated strictly by the overall utility of a single stakeholder, as is often the case of more mainstream recommender system applications. In this article, we focus our discussion on the challenges of multistakeholder evaluation of recommender systems. We bring attention to the different aspects involved---from the range of stakeholders involved (including but not limited to providers and consumers) to the values and specific goals of each relevant stakeholder. We discuss how to move from theoretical principles to practical implementation, providing specific use case examples. Finally, we outline open research directions for the RecSys community to explore. We aim to provide guidance to researchers and practitioners about incorporating these complex and domain-dependent issues of evaluation in the course of designing, developing, and researching applications with multistakeholder aspects. 
\end{abstract}

\maketitle

Keywords: recommender systems, evaluation, multistakeholder issues

\include{ijhcs_article_v2}

\bibliography{ms-eval}

\end{document}

%% file: ijhcs_article_v2.tex
\section{Introduction}
\label{sec:Intro}
Understanding the performance of any machine learning system requires evaluation. Research in recommender systems has focused almost exclusively on evaluating the experience of the end user receiving recommendations, even though recent research highlights the crucial role that recommender systems play in larger ecosystems of commerce and media distribution \cite{abdollahpouri2020multistakeholder,boutilier2024recommender}. To encompass the larger context of a complex recommender ecosystem requires taking a broader view of evaluation (although end users will always be important). This brings us to the topic of multistakeholder evaluation as defined in \cite{abdollahpouri2020multistakeholder}:

\begin{quote}``A \textbf{multistakeholder evaluation} \index{multistakeholder evaluation} is one in which the quality of recommendations is assessed across multiple groups of stakeholders.''
\end{quote}

The aim of this article is to provide an introduction to the topic of multistakeholder evaluation of recommender systems. We provide (i) an overview of the types of recommendation stakeholders that may be considered in conducting such evaluations, (ii) a discussion of the considerations and values that enter into developing measures that capture outcomes of interest for a diversity of stakeholders, (iii) an outline of a methodology for developing and applying multistakeholder evaluation, and (iv) three examples of different multistakeholder scenarios including derivations of evaluation metrics for different stakeholder groups in these scenarios. While it goes without saying that a system designed with multiple stakeholders in mind will need to be evaluated relative to the perspectives of such stakeholders, multistakeholder evaluation does not necessarily imply that the recommender system in question accounts for multiple stakeholders in its design or optimization objectives. One may wish to audit the performance of the system relative to different stakeholders interests without necessarily feeding all such measures back to the algorithm. 

This work derives from a Dagstuhl seminar titled \textit{Evaluation Perspectives of Recommender Systems: Driving Research and Education}~\cite{zangerle_et_al:2024}.\footnote{\url{https://www.dagstuhl.de/en/seminars/seminar-calendar/seminar-details/24211}, May 2024}
This article reflects the perspectives of a sub-group of the  researchers in attendance focused on the multistakeholder evaluation of recommender systems~\cite{burke2024multistakeholder}. These researchers individually contribute knowledge in multistakeholder methodologies, their evaluation, as well as application to several domains (music, HR, and education). In addition, the cumulative experience of these researchers points to gaps in the multistakeholder evaluation for recommender systems. Notably, we argue that the research to date does not have a clear method for developing evaluations that reflect stakeholder impact or the wider range of possible impacts. In this article, we make a case for revisiting value-sensitive design (c.f., \cite{friedman1996value}) and indicate necessary steps for developing evaluation protocols, in addition to settling on metrics, and handling trade-offs in stakeholder preferences. Finally, we argue that multistakeholder evaluation necessarily reflects (often implicit) values (c.f., \cite{starke2024normalize}), and that impactful evaluation requires the identification not only of a wider range of stakeholders, but also the specific characteristics of a task such as: application specificity, context specificity, and institutional sensitivity.

Evaluating a methodology aimed at the development of software applications usually goes through the definition of three main components \cite{10.1145/251880.251912}: (i) a list of feature requirements, (ii) a method of scoring the features in the methodologies targeted, and (iii) a set of guidelines for applying the evaluation framework. Among these components, a first-class role is played by the identification of relevant features to assess the evaluation. A key element for the evaluation is the identification of similarities and differences with other methodologies as well as to set a common reference. This latter can always be used to set a comparative evaluation of different methodologies as well as of different choices within the same methodology. In the multistakeholder scenario we target, a common reference baseline can always be set as the "classical" consumer-only one. 

In this work, we provide the reader with all the elements to evaluate the adopted methodology with reference to the components mentioned before.
\begin{itemize}
    \item Given the targeted stakeholders (a classification is provided in Section \ref{sec:stakeholders}), for each of them the system designer defines a set of features that follow clear values and goals as highlighted in Section \ref{sec:multis:values}.
    \item The selected features, referring to different stakeholders, can be measured and scored, as discussed in Section \ref{sec:methodology:metrics}, and eventually aggregated (see Section \ref{sec:aggregation}) to provide an overall holistic evaluation of the selected features/values.  
    \item A set of practical guidelines are then provided in Section \ref{sec:methodology:guidelines} for conducting multistakeholder evaluation. 
\end{itemize}

In order to support the application of the multistakeholder evaluation, in Section \ref{sec:multis:examples} we show how to apply the above mentioned guidelines with the introduction of some example scenarios and metrics. 
Finally, in Section \ref{sec:challenges} we describe potential challenges in multistakeholder recommendation.

\subsection{Stakeholder types}\label{sec:stakeholders}
\begin{figure}[tbh]
    \centering
    \includegraphics[width=1\linewidth]{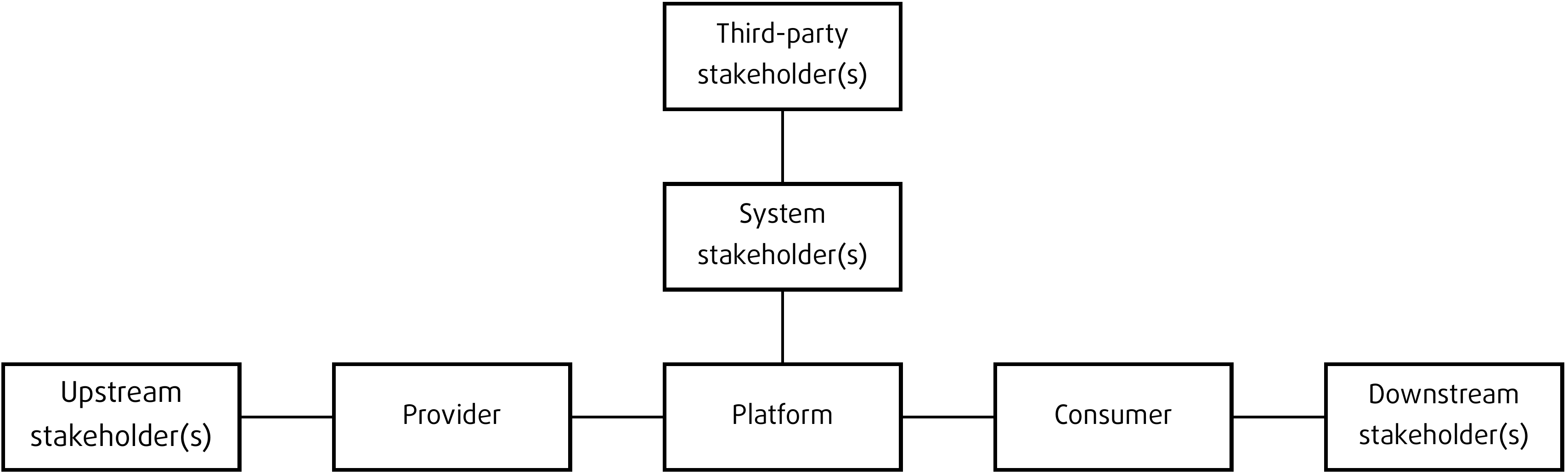}
    \caption{A multistakeholder view of a recommendation ecosystem}
    \label{fig:ms-overview}
\end{figure}

The variety of stakeholders forming a recommendation ecosystem is suggested in Figure~\ref{fig:ms-overview} and defined here, extending the terminology from \cite{abdollahpouri2017recommender,abdollahpouri2020multistakeholder}:

\begin{itemize}
    \item[Recommendation \textbf{consumers}] are the traditional recommender system end users to whom recommendations are delivered and to which typical forms of recommender system evaluation are oriented.
    
    \item[Item \textbf{providers}] form the general class of individuals or entities who create or otherwise stand behind the items being recommended. 
    
    \item[\textbf{Upstream} stakeholders] are those potentially impacted by the recommender system through the provider side of the interaction, but who are not direct contributors of items. For example, in a music streaming recommender, a songwriter may receive royalties based on songs that are played, but it is the musical artist's performance of the respective song that is the item actually being recommended and listened to. 
    
    \item[\textbf{Downstream} stakeholders] are those who are impacted by choices that recommendation consumers make, by interacting with chosen items or being impacted by the use or consumption of recommended items. For example, in a recommender system that suggests children's books to teachers, the children who ultimately get the books (and their parents) are downstream stakeholders from the teachers who are the consumers, receiving recommendations from the system \cite{ekstrand2018retrieving,ekstrand2023seeking}.
    
    \item[\textbf{System} stakeholder] is intended to stand in for the organization creating and operating the recommendation platform itself. A recommender system is implemented and maintained for a reason, in support of an organization's goals. An e-commerce site may implement a recommender system to achieve goals around user satisfaction and retention, quite related to the goals of consumers, but they may also have economic goals (increasing average purchase basket size, for example) that are not necessarily shared by the consumers or providers.
    
    \item[\textbf{Third-party} stakeholders] are those individuals or groups who do not have direct interaction with the system that nonetheless have an interest in, or are impacted by, its operation. For example, in a domain such as job recommendation, government agencies charged with ensuring non-discrimination in hiring practices may be considered stakeholders whose requirements are legally binding on the platform operator. 
\end{itemize}

As we have noted, the vast majority of recommender systems research focuses its evaluation only on the perspective of recommendation consumers. However, in most applications, numerous stakeholders are involved in the upstream and downstream parts of the provisioning, recommending, and consumption process. Here, we illustrate this complexity using a (hypothetical) music streaming application as an example---additional examples from other application areas are described in Section~\ref{sec:multis:examples}. 

\begin{figure}[tbh]
    \centering
    \includegraphics[width=1\linewidth]{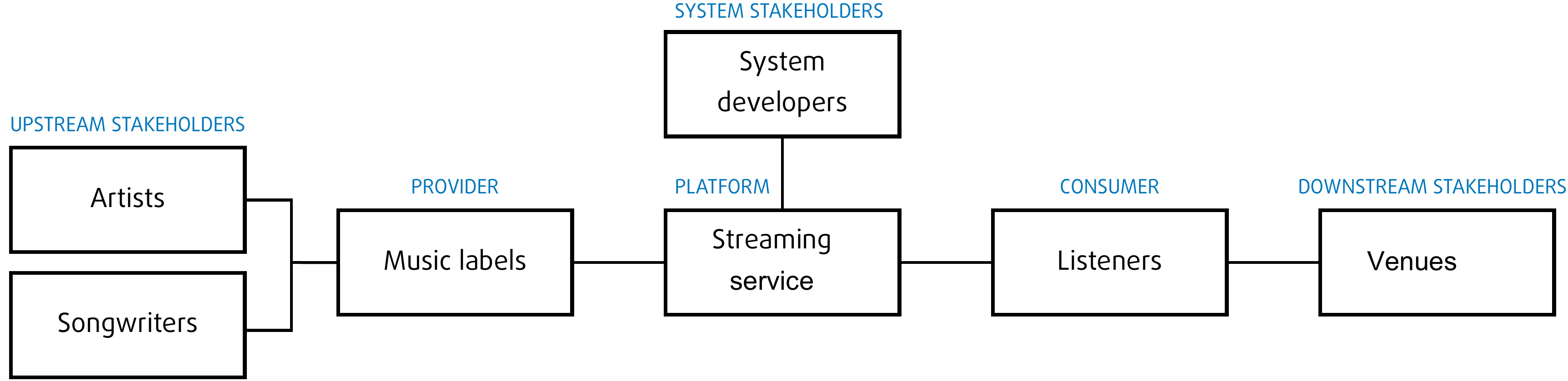}
    \caption{Stakeholder configuration for the music streaming example.}
    \label{fig:ms-example-music}
\end{figure}

Figure \ref{fig:ms-example-music} shows the different stakeholders involved in the system, with songwriters, artists, and label companies on the content production and provisioning side. The platform (recommender system) plays the role of mediating between upstream and downstream stakeholders. On the downstream side, consumers are the first-line stakeholders, but others may also be affected by the recommendations, e.g., owners of concert venues where recommended artists might appear.

Stakeholders pursue specific goals that are driven by values (see Section \ref{sec:multis:values}) meant as generic concepts helping an individual to choose the best actions or behaviors~\cite{rokeach1973nature,schwartz2012overview}. While values are generic concepts and may apply across a wide range of application domains, goals can be seen as intermediate-level objectives that are operationalizations of, for example, a generic human- or business-centric value. Each goal can be assessed by different measures, which may be captured using a variety of concrete measurement methods and metrics \cite{ekstrand2024not}. For the music streaming example, we present sample goals and measures in Table~\ref{tab:stakeholders-music}; these are described in greater detail in rest of the article. 

Unsurprisingly, the goals of different stakeholders may compete with each other. A system that is seeking to satisfy competing stakeholders will need to balance stakeholder goals in the recommendation process. In this example, conflicting goals may be that system operators want to increase the monetary benefit by favoring popular artists and songs, which might negatively affect the visibility of long-tail artists who want to build an audience.\footnote{We stress that all examples in this discussion are hypothetical and may or may not represent actual stakeholder configurations or goals. For additional perspectives on multi-objective recommendation in music recommendation, see \cite{unger2021deep}.}

\begin{table}[tbh]
\resizebox{1\textwidth}{!}{%
\begin{tabular}{|l||c|c|c|c|c|}
\hline
\textbf{}            & \textbf{Upstream}   & \textbf{Provider} & \textbf{System}   & \textbf{Consumer} & \textbf{Downstream} \\ \hline \hline
\textbf{Stakeholder} & \begin{tabular}[c]{@{}c@{}}Artist / \\ Songwriter\end{tabular} & \begin{tabular}[c]{@{}c@{}}Music \\ Label\end{tabular}      & \begin{tabular}[c]{@{}c@{}}Streaming \\ Service\end{tabular} & Listener          & \begin{tabular}[c]{@{}c@{}}Concert \\ Venue\end{tabular}         \\ \hline
\textbf{Goals} &
  \begin{tabular}[c]{@{}c@{}}Monetary \\reward, \\ Reputation and \\ recognition\end{tabular} &
  \begin{tabular}[c]{@{}c@{}}Monetary reward, \\ Market \\development, \\ Product planning\end{tabular} &
  \begin{tabular}[c]{@{}c@{}}Monetary \\reward, \\ Customer loyalty\end{tabular} &
  \begin{tabular}[c]{@{}c@{}}Enjoyment, \\Wellbeing, \\ Personal \\development\end{tabular} &
  \begin{tabular}[c]{@{}c@{}}Monetary reward, \\ Market \\development, \\ Schedule planning\end{tabular} \\ \hline
\textbf{Measures} &
  \begin{tabular}[c]{@{}c@{}}Revenue, Royalty, \\ Exposure, User \\ feedback, Playlist \\ inclusion\end{tabular} &
  \begin{tabular}[c]{@{}c@{}}Revenue, Exposure, \\ Consumption\\ trends, \\ User feedback\end{tabular} &
  \begin{tabular}[c]{@{}c@{}}Revenue, \\Customer \\ retention, User \\ feedback\end{tabular} &
  \begin{tabular}[c]{@{}c@{}}Ratings, Reviews, \\ Music knowledge, \\ Sharing\end{tabular} &
  \begin{tabular}[c]{@{}c@{}}Ticket \\\& Merchandise \\ Sales, Concertgoer \\feedback \end{tabular} \\ \hline
\end{tabular}
}
\caption{Sample stakeholder goals and measures for the music streaming example.}
\label{tab:stakeholders-music}
\end{table}

\subsection{Scope}
The topic of evaluation touches on many aspects of recommender systems design, implementation and maintenance, more than can be encompassed in a single article. Here, we focus on the problem of principled derivation of evaluation metrics based on an existing system design. We aim to provide guidance to researchers, practitioners, and others who seek to incorporate multistakeholder evaluation into their analysis of recommender system properties and outcomes.

We do not focus on the relationship between evaluation and system design itself, assuming that a system already exists, designed to meet a particular information need, embedded within a particular stakeholder ecosystem. System stakeholders would already have formulated evaluation metrics intended to capture the value that they expect from the recommender system and would have optimized the system to meet those objectives. System designers may or may not have incorporated diverse stakeholder perspectives in their work, but regardless of the history of design decisions, the impact of the system relative to different stakeholders can still be evaluated. 

If developers take a multistakeholder perspective in developing a new system, they would need to engage in many of the analyses outlined in this article to understand how to evaluate the system. Simultaneously, they would have to consider how the recommendation task is defined and how the spectrum of evaluation criteria can be incorporated into the optimization of recommendation models and the delivery of recommendations.  

Another topic that we do not address is the design of evaluation methodologies. It is certainly the case that some outcomes (for example, user opinion about the qualities of recommendation lists) can only be measured through surveys or other user studies, whereas other properties (for example, the number of items of a particular type appearing in recommendation lists) can be measured from system outcomes. System properties can be measured in online and offline ways. Readers are referred to the extensive literature on recommender systems evaluation (particularly the overview and surveys in  \cite{zangerle2022evaluating,gunawardana2022evaluating,knijnenburg2012explaining}). However, it should be noted that these methodologies are almost exclusively aimed at measuring user-oriented outcomes, with limited research available on evaluation methodologies specifically tailored for other stakeholder outcomes.

\subsection{Challenges}

Even within the scope that we have chosen for our study, researchers and practitioners face several key challenges that go beyond those typically encountered in recommender systems research and of which they should be aware. 

\begin{itemize}
    \item \textbf{Application specificity}: Recommender systems research is, in general, highly domain specific. This specificity is even more pronounced when the larger ecosystem is considered. As our examples make clear (see Section \ref{sec:multis:examples}), different recommendation applications have different stakeholder configurations and different types of benefits that stakeholders may gain. Even across recommender applications for which outputs are superficially similar (for example, music playlists), stakeholders may occupy different niches and require different analyses. 

    \item \textbf{Access to data}: Typical recommendation datasets have little to no information about non-consumer stakeholders, so it is difficult to understand what are realistic calculations of, for example, revenue distribution among item providers. Additional investigation will typically be required to gather the data needed to design effective multistakeholder evaluations.

    \item \textbf{Context specificity}: Different legal regimes and cultural differences may impose different regulatory requirements on recommender systems, and it is therefore difficult to formulate constraints from third-party stakeholders in a general way. 

    \item \textbf{Institutional sensitivity}: There is a strong tradition in research and writing about recommender systems to emphasize the primacy of consumer-side outcomes. This is evident in user interface language of the systems, e.g., via the use of ``Recommended for you'' and similar labels. Recommendation platforms are often reluctant to publicize or discuss multistakeholder aspects of their systems, even though incorporating such considerations is standard practice.\footnote{As one example, we note that, buried at the bottom of its page on recommendations (\url{https://www.spotify.com/us/safetyandprivacy/understanding-recommendations}), Spotify states the following ``Spotify prioritizes listener satisfaction when recommending content. In some cases, commercial considerations, such as the cost of content or whether we can monetize it, may influence our recommendations.''  Such transparency is rare in the industry.}

    \item \textbf{Adversarial aspects}: Recommendation platforms may actively discourage providers in particular from acquiring knowledge about the platform that might enable strategic activity: for example, misrepresenting their items to gain algorithmic favor. There is no doubt that providers are sometimes incentivized to do this, as the history of search engine spam attests. Although adversarial considerations do not limit a platform's ability to conduct internal multistakeholder evaluation, it is an open research question how to design evaluation metrics and disclosure protocols such that outcomes can be shared with providers without enabling adversarial behavior. 

\end{itemize}

\subsection{Related Work}
Multistakeholder recommendation has long been a subject of study in the field although the term itself was not coined until 2017 \cite{abdollahpouri2017recommender}. The (pre)history of multistakeholder recommendation is detailed by \citet{abdollahpouri2020multistakeholder}. In these earlier works, researchers defined evaluation approaches deemed appropriate to their particular task, measuring the outcomes important to the aspects of recommendation under consideration. For example, early work by \citet{chen2008developing} combined profitability with standard measures of recall and precision in evaluating a product recommender. However, because multistakeholder recommendation was not conceptualized as a research area, connections between these research areas and a general theory of multistakeholder evaluation in recommender systems did not materialize. 

\citet{abdollahpouri2020multistakeholder} call out multistakeholder evaluation specifically as a research topic in their survey. They specifically formulate a number of possible provider metrics. However, unlike our approach here, this work is not grounded in value-centric consideration of stakeholder perspectives. They call for practitioners to formulate metrics from the point of view of different stakeholders but do not consider making use of stakeholder involvement in the process. Similarly, in the area of fairness-aware recommendation, a subtype of multistakeholder recommendation, \citet{ekstrand2022fairness} survey a wide variety of fairness metrics that have been used in the literature. However, these metrics are generally not the product of stakeholder consultation. Some steps in this direction are taken in \cite{smith2023many} and \cite{smith2024recommendme} where stakeholder consultation in the form of qualitative interviews is used to derive fairness concerns and possible operationalizations. Multistakeholder aspects of machine learning fairness in industry practice more generally are explored in \cite{smith2025pragmatic}. This work finds that practitioners have little leeway to conduct broad stakeholder consultation in most circumstances, even when they believe this may be valuable. Practitioners report that they have little to go on when trying to understand and operationalize objectives representing different stakeholder interests.

In general, the Dagstuhl seminar participants found that there was insufficient guidance available on how to design and conduct multistakeholder evaluation of recommender systems. In particular, we believe that evaluation must be rooted in understanding what stakeholders value in a system and that this step of analysis is treated implicitly or not at all in most prior work.  

\section{Values}
\label{sec:multis:values} 
\citet{jannach2022value} argue that, in the ideal case, recommender systems would ``\textit{create value in parallel for all involved stakeholders}''. At the same time, it is unavoidable for competing goals to arise, since direct and indirect stakeholders, including the system itself, may have differing perspectives. In this case, to \textit{evaluate} the value created for those involved, we argue that it is imperative to go back to a fundamental and normative question and one that is rarely asked according to \citet{jannach2012value}: ``\textit{What
is a good recommendation (in a given context)?}''

To answer this complex question, we posit that one first must look into the values each stakeholder aims for in the recommendation process. The concept of value has been discussed in the recommender systems literature from multiple perspectives \cite{jannach2016recommendations,torkamaan2024role,abdollahpouri2020multistakeholder,de2023systematic,stray2024building,ghanem2022balancing,murgia2019seven}. Perhaps the most prominent are those referring to the business side of the equation (provider-or system-centered) or the user side (consumer-centered), i.e., the utility of the ultimate consumer. From a more human perspective, values concerning individuals directly or indirectly served by recommender systems~\cite{stray2021you,stray2024building,konstan2021human,ekstrand2022recommender}, and those with societal implications~\cite{fabbri2023social,milano2020recommender,milano2021ethical,gonccalves2024societal,stray2024building} have also been discussed, such as the level of social utility that it achieves, the social rights that it strengthens or harms, and the existing social norms that it alters or reinforces.  The concept of values in domain-specific implementations of recommender systems has also been investigated, e.g., news recommender systems~\cite{bauer2024values} or recommender systems for humanities and historical research~\cite{atzenhofer2024value}. Although these works analyzed a specific domain for recommender systems, we find that many of the identified values overlap and that both more business-oriented and more human-oriented values are discussed. 

In the context of this work, we refer to value as standards or criteria that help an individual to select and evaluate actions or behaviors~\cite{rokeach1973nature,schwartz2012overview}. With that in mind, for multistakeholder recommender systems, the term value might refer to standards (or even a set of standards) a stakeholder expects or imposes on the recommendation process. The significance of values in system design has been highlighted within the field of human-computer interaction through the development of value-sensitive design processes. We point the interested reader who seeks a deeper understanding on value-sensitive design for information systems to the literature in the field~\cite{miller2007value,friedman2013value,friedman2002value,borning2012next}. 

In multistakeholder recommender systems, values must be considered when evaluating the `goodness' not just of a recommendation in isolation, but of the stakeholders that are part of the entire process within the specific contexts and domains in which the systems are deployed. Thus, these systems can be conceptualized as complex socio-technical systems in which algorithmic processes are embedded within broader structures of negotiation, deliberation, and governance. Particularly in public contexts, values are not merely elicited from stakeholders but are shaped through deliberative processes that mediate trade-offs and articulate shared goals. This perspective highlights the importance of viewing value alignment not just as a technical optimization problem, but as a process of participatory design. Insights from public policy and political science, such as theories of deliberative democracy, stakeholder governance, and collaborative planning, can therefore enrich the value elicitation processes of multistakeholder recommender systems by grounding them in frameworks that acknowledge power asymmetries, legitimacy, and fairness. We therefore, also point the interested reader to the rich body of literature in this realm~\citep[e.g.,][]{dryzek2002deliberative,hajer2003deliberative,stilgoe2020developing}. 

Value-associated considerations in technology have in recent years also become subject of political discourse and policy making. The European Union, for instance, has issued legal regulations that aim at safeguarding against adverse effects of digital services in general and, specifically, AI-based systems. The Digital Services Act \footnote{\url{https://eur-lex.europa.eu/legal-content/EN/TXT/?uri=CELEX3A32022R2065}} (DSA) addresses a wide range of online platforms, generally setting the goal of ``ensuring a safe, predictable and trustworthy online environment''. The DSA repeatedly spotlights recommender systems, requesting, inter alia,  that ``recipients of their service are appropriately informed about how recommender systems impact the way information is displayed, and can influence how information is presented to them''. Providers should also clearly present the parameters, at least the most important criteria, to let users understand how the information is prioritized for them. The impact of the DSA on multistakeholder recommender systems has already become subject of scientific discourse (see, e.g.,~\cite{naudts_right_2025}). 

Specific regulations were recently also adopted for AI-based systems \footnote{\url{https://eur-lex.europa.eu/eli/reg/2024/1689/oj/eng}} (the EU AI Act). The approach taken there is risk-based, classifying applications into several risk classes. While many recommender systems (as one relevant class of AI-based systems) can be assumed to involve only acceptable risk for stakeholders, e.g., in music streaming, other domains addressing, e.g., the job market or medical applications may indeed create substantial risks for some stakholder groups~\cite{kowald2024establishing}. Identifying such risks should therefore become integral part of the value elicitation process for stakeholder groups that may be critically affected. Both the DSA and the AI Act are relevant for different stakeholder groups, but do not specifically address multistakeholder issues.

In the rest of this section, we review seminal literature that provides background on the concept of value from different perspectives and its connection to recommender systems. Along the way, we highlight the most common values to consider when evaluating multistakeholder recommender systems. It is worth noting that the values we mention are not meant to be an exhaustive list. Instead, they serve as a starting point to encourage reflection among researchers and practitioners to move beyond the more typical `producer versus consumer' perspective and consider a myriad of factors to (simultaneously) account for when evaluating multistakeholder recommender systems.

\subsection{Economic and Business-Related Values}

When addressing values in the context of multistakeholder recommender system evaluation, economic and business-related values are often considered, especially for providers and system operators. 
De Biasio et al. \cite{de2023systematic} provide a systematic review of value-aware recommender systems, introducing value primarily as an economic concept leading to \textbf{monetary reward} (i.e., profit and revenue). They distinguish several aspects that inform the value of monetary reward reflective of a business and economic view, including use value (e.g., increasing revenue by providing useful recommendations), estimated value (related to attractiveness and desirability, such as having a comprehensive music catalog to create recommendations from), cost value (e.g., the economic resources required to distribute a music album to the music streaming platform), and exchange value (the change in value over time, e.g., increase in a music artist's recognition and popularity on the platform due to effective recommendations). 

From this perspective, we also observe values related to \textbf{user perception} and \textbf{customer loyalty}, which are crucial from both a business and economic perspective. These values often relate to \textit{``the concepts of quality and personalization, experience and trust, features, and benefits''}~\cite{de2023systematic}. For example, in the music industry, a platform that provides highly personalized playlists based on users' listening history can significantly enhance user satisfaction. This personalization not only helps users discover new music that aligns with their preferences but also fosters a sense of trust and loyalty towards the platform. Users are more likely to stay subscribed and recommend the service to others if they consistently experience high-quality, relevant recommendations. 

The authors in \cite{debiasio2023economic} highlight that recommender systems typically serve an organization's economic values. Besides profit and revenue (i.e., monetary rewards), this might be related to \textbf{growth and market development}. For example, music streaming platforms aim to generate profit and attract new users by offering social features like joint playlist creation, which benefit users when their peers are also on the platform. 
Furthermore, the authors characterize economic recommender systems as systems that exploit ``\textit{price and profit information and related concepts from marketing and economics to directly optimize an organization's profitability.}'' 
 \citet{jannach2016recommendations} identify strategic perspectives for both consumers and providers. For consumers, personal utility includes happiness, satisfaction, knowledge, and entertainment. For providers, organizational utility encompasses profit, revenue and growth. In addition, other values, such as \textbf{changing user behavior to create demand}, might be relevant. For example, a music streaming platform might recommend emerging artists or newly released tracks to users, encouraging them to explore and adopt new music preferences, thereby creating demand for content that the platform can monetize.

\citet{jannach2022value} examine the theory of business models in e-commerce recommender systems and identify the following value-driving aspects: \textbf{efficiency} (e.g., the exposure of music artists in recommendation lists or the number of clicks on recommended music tracks), \textbf{complementarities} (e.g., creating value through synergies by combining different item types like recommending merchandise articles along with track recommendations of a specific music artist), \textbf{lock-in and churn prevention} (e.g., retaining subscribed users by providing helpful recommendations), and \textbf{novelty and product planning} (e.g., finding new fans through recommendations to users who might like an artist's music or getting inspired to create new music album). 

In addition to immediate financial outcomes, recommender systems can enhance a platform's \textbf{brand equity}, which refers to the value added to a product or service by its brand as perceived by consumers. Brand equity is built through positive user experiences that increase user satisfaction and loyalty, both critical components of brand equity \cite{aaker1996measuring,washburn2002measuring}. 
\citet{jannach2019measuring} emphasize that well-crafted recommendations enable platforms to differentiate themselves in competitive markets, thereby strengthening their reputation.  

Similarly, \citet{maslowska2022role} suggest that when recommendations align with users' personal goals, they not only encourage engagement but also create a positive spillover effect on the platform's brand, enhancing overall trust and loyalty. 
Recommender systems also open up \textbf{cross-selling and up-selling} opportunities in ways that reinforce brand value, such as suggesting premium products or exclusive experiences that align with the brand’s identity (e.g., premium subscriptions or concert tickets). These strategies, which refer to offering related products or higher-end options to users, help increase sales volume and raise average revenue per user \cite{jannach2019measuring,kubiak2010cross,yeniaras2024cross}.
Finally, platforms can also use user interaction data from recommender systems to \textbf{develop personalized marketing strategies} while maintaining high privacy standards, a practice that reinforces consumer trust and encourages long-term engagement \cite{maslowska2022role}.

\subsection{Societal and Human-Centric Values}
Beyond economic and business values, societal and human-centric values, which cover other important aspects, are also crucial for businesses and platforms. Societal and human-centric values for stakeholders in recommender systems focus on ensuring that these systems operate in ways that prioritize humans individually and society as a whole. We find that there are 4 themes of societal and human-centric values for stakeholders in recommender systems that are relevant in the light of evaluation: (i) usefulness, (ii) well-being, (iii) legal and human rights, and (iv) public discourse and safety~\cite{stray2024building,torkamaan2024role,ekstrand2022recommender}. 

\textbf{Usefulness and enjoyment} means that recommendations should meet the needs and expectations of its stakeholders effectively and efficiently~\cite{knijnenburg2012explaining}. For example, in the case of a music recommender system, users should be able, via the recommender system, to discover new music that they might enjoy and match their tastes. At the same time, usefulness refers to the recommender system's ability to help music artists get their outputs recommended to potentially interested listeners. \textbf{Control and privacy} are closely related values that pertains to the degree of influence and customization stakeholders might have over the recommendations that are generated.  This includes privacy aspects in a way that users might want to control the amount of their (music) preference data that is shared with the recommender system~\cite{muellner2021robustness,mullner2023reuseknn,stray2024building}.  

\textbf{Well-being} refers to the extent to which a recommender system supports and enhances users' psychological, emotional, and social health rather than simply optimizing engagement or consumption~\cite{ekstrand2022recommender}. In the case of a music recommender system, this means that recommendations should influence the experience with the music streaming platform positively, e.g., provide music recommendations to help listeners relax or relieve stress~\cite{knees2019user}. In this respect, well-being is related to emotional, mental, psychological, physical, and social health. Other related values are \textbf{connection, community and social bonding}, e.g., to enable users to connect with like-minded music listeners or to enable music artists to contribute their outputs to a specific community. Thus, also \textbf{reputation, recognition and acknowledgement} might be valuable for some stakeholders, e.g., to support music artists in getting their contributions recognized by music listeners~\cite{milano2021ethical}. \textbf{Personal growth and development} might also be values contributing to well-being in the sense that, e.g., music recommendations could help people explore new music styles and genres, supporting exploration and self-discovery~\cite{celma2008hits}.  

Concerning legal and human rights, \textbf{fairness} might be an important value for stakeholders of a recommender system at evaluation time. For example, the music streaming platform should aim to provide meaningful recommendations to all user groups, independent of, e.g., their musical taste, demographic characteristics, or inclination towards popularity~\cite{ekstrand2018all,lesota2021analyzing,kowald2022popularity,dinnissen2022fairness}. Additionally, the music recommender system should aim to treat music artists fairly and, in that sense, include novel or (less popular) ``niche'' artists in the recommendation lists when applicable~\cite{kowald2020unfairness,sonboli2022multisided}.

Fairness can be related to diversity when the goal is to ensure, for example, that diverse articles and styles are represented in recommendation outputs. Diversity may also have listener-oriented benefits, e.g., help music listeners explore artists that might be new to them \cite{porcaro2021diversity,duricic2021my}. A recommender system might seek to support \textbf{freedom of expression} as well as \textbf{accessibility and inclusiveness} by allowing, e.g., music artists to promote their content independent of the genre or popularity of their music~\cite{bell2022assessing,ramadhani2024freedom}. At the same time, recommender systems should enable users to access the content that they like and enjoy, even when their taste does not match the majority of music listeners~\cite{ferraro2019music,kowald2021support}. \textbf{Transparency and trustworthiness} might also be an important value for all stakeholders of a recommender system. For instance, music artists might be interested in why they are ranked at a specific position and music listeners might be interested in why a specific artist was recommended to them~\cite{moscati2023integrating,sinha2002role}.  

Values in the area of public discourse and safety are related to a multitude of societal and human-centric aspects. Here, \textbf{societal benefit} goes beyond the satisfaction of individual stakeholders. As an example, a music streaming platform might be interested in fostering cultural enrichment by recommending a diverse set of music~\cite{vargas2011rank}. This is related to the value of \textbf{tradition and history}, for instance, by recommending local and traditional music, which might be hard to find without the recommender system \cite{ferraro2021fair,lesota2024oh}. The \textbf{environmental sustainability} might also be an important value for some recommender systems stakeholders. This may involve implementing energy-efficient recommendation models within the platforms, or promoting local music artists whose concerts offer the opportunity for attendance without requiring extensive travel~\cite{merinov2023sustainability}. Finally, \textbf{safety} is concerned with users not being exposed to recommendations of disturbing ethically questionable, or age-inappropriate content. In the case of music recommendations, this could refer to music tracks with offensive lyrical content~\cite{messner2007hardest,council2009impact}.

\subsection{Values in Practice}
As we mentioned earlier, the concept of value can be perceived as abstract. We can think of values as high-level, somewhat theoretical, and qualitative objectives that, in practice, must be refined for quantification \cite{ekstrand2024not}. As mentioned by \citet{beattie2022challenges}, this process involves scoping ``what exactly needs to be measured." This inherently complex challenge becomes even more obvious in the context of multistakeholder recommender systems, where different stakeholders--such as producers and consumers--may prioritize distinct values, each requiring unique and/or conflicting evaluation approaches \cite{beattie2022challenges}.

Nevertheless, when it comes to evaluation of multistakeholder recommender systems, we must find a way to quantify values. If the aim is to determine the `goodness' for all involved; there is a need for "methods for measuring how well a system is adhering to, promoting, or facilitating these values over time" \cite{stray2024building}. 
Translating theoretical values into concrete, measurable concepts requires tailored quantitative and/or qualitative methodologies specific to each multistakeholder use case  \cite{ekstrand2024not}. 

As an initial reference, we can look at the work by \citet{wu2022multi} focused on the fairness value. Although limited to consumer-sided and producer-sided fairness--two of the many stakeholders we mention in this narrative--their proposed framework for optimization and the metrics considered for overall evaluation, offer a starting point for exploration that could be extended to multiple stakeholders concurrently. The recent synthesis by \citet{stray2024building}, focused on human and societal-related values, provides valuable insights into how to translate these values into measurable metrics for assessment. However, much like we discuss earlier in this section, the authors emphasize that in a multistakeholder context, there is neither a definitive list of values nor a single method for quantifying them. Instead, they present possible interpretations of values within the context of (multistakeholder) recommender systems and suggest potential indicators for evaluating whether a system upholds or supports these values (for more details, see the Appendix in \cite{stray2024building}).  

Ultimately, there is no universal approach or single measure for evaluation, as the process depends on the stakeholders involved, the focus of the recommendation application, and many other factors. As an exhaustive checklist for holistically evaluating multistakeholder recommender systems is not feasible, we provide in Section \ref{ms-methodology} general guidance on navigating this evaluation process.

\section{Methodology}
\label{ms-methodology}

As we previously noted, evaluating recommender systems is a contextually situated problem: different domains, recommendation tasks, and contexts require specific metrics and evaluation setups tailored to that specific recommendation scenario. 
Multistakeholder evaluation, where the perspectives of other stakeholders are taken into account in addition to that of the consumer, only increases the potential complexity of the evaluation. The complexity of multistakeholder evaluation is demonstrated by the richness and variety of the examples described in Section \ref{sec:multis:examples}. 
As a result of this complexity, prescribing exactly which methods to use in which order is impractical. Instead, we attempt to describe best meta-practices for conducting successful multistakeholder evaluation in this section, divided into different stages. We consider this process to be iterative, as findings in a later stage can necessitate returning to an earlier stage, for instance, when learning of a new relevant stakeholder to include or when value shifts occur in one or more stakeholders. These stages are

\begin{itemize}
    \item Stakeholder identification
    \item Value articulation
    \item Operationalization of values as metrics
    \item (if needed) Aggregation of metrics into an overall system objective
\end{itemize}

Recall that, in our discussion, we assume that we seek to evaluate an existing recommender system, one that has already been developed to provide a particular recommendation function. Of course, planning for a system's evaluation should be part of its development: stakeholder consultation would naturally be prioritized in the design and implementation of a multistakeholder recommender system. 

\subsection{Stakeholders}\label{sub:methodology:stakeholders} 
The cornerstone of multistakeholder evaluation is identifying the relevant stakeholders that will be affected by or affect the recommendation process in some way, as shown in Figure~\ref{fig:ms-overview}.
%
The core parties in any multistakeholder evaluation are the consumers, providers, and the system stakeholders behind the recommendation platform. A sensible first step is to engage with the system stakeholders and gauge their understanding of whom they are recommending to (= consumers) and where the items being recommended come from (= providers). System stakeholders, by virtue of their central role, are also most likely to have the greatest awareness of potential third-party stakeholders whose decisions may impact the operation of the recommendation platform. Commonly, third-party stakeholders would involve regulatory bodies and institutions; here, the system stakeholder's legal department could help identify relevant regulations (e.g., related to consumer protection) and the right parties to reach out to.
Finally, depending on the recommendation scenario, system stakeholders may also be helpful in identifying relevant upstream and downstream stakeholders. 

Consumers (or users) have historically played (and will continue to play) a central role in recommender systems evaluation. As a result, a common next step would be profiling the consumer stakeholder and the different subgroups this stakeholder category may represent. In addition to interviews with the system stakeholders, any existing market or user research on the user base of the recommendation platform could serve as a valuable foundation for identifying representative subgroups within this user base. A literature review aimed at identifying similar or related recommendation scenarios could also help identify different user groups, especially groups that may be underrepresented in the market research for whatever reason.
The system stakeholder should be able to facilitate access to these subgroups, for instance through user research panels, surveys on the website, or customer mailing lists. It is important to recruit a diverse and representative sample of consumers to represent the customer stakeholder and ensure all voices are heard in the evaluation process. Customers should be interviewed or surveyed about what values matter to them in this recommendation scenario (and their relative importance), what goals they have, and how and when they envision using the recommender system.
If representative, the principle of saturation could be useful in guiding the sample size required: if additional participants do not reveal any new values, goals, or usage scenarios, then the sample should be representative of the customer stakeholder. Consumers are also a valuable source for identifying possible downstream stakeholders that are worth including in the evaluation process.

The item provider(s) are the general class of individuals or entities who create or otherwise stand behind items being recommended. Historically, they have perhaps been less well represented in recommender systems evaluation, but they play an essential role in multistakeholder evaluation. 
The number of different individuals or entities that make up the provider stakeholder role may vary greatly between recommendation scenarios: in some cases, only a handful of entities may be providing the items to be recommended, whereas in others they may be as numerous as consumers. 
Similar to the customer stakeholder, the system stakeholders should be able to facilitate access to the provider stakeholders and help identify which of them carry the biggest weight, without losing sight of the relevant minority providers. Providers are the most valuable source for identifying possible upstream stakeholders that are worth including in the evaluation process.
Again, it is important here to recruit a diverse set of representatives for this stakeholder group to ensure that their needs, values, and goals are all met in the evaluation process.

One outcome of interviewing the consumer, provider, and system stakeholders should be the identification of any relevant upstream and downstream stakeholders. This could be supplemented with additional stakeholders identified through a literature review aimed at identifying similar or related recommendation scenarios.

Each stakeholder group should be involved in the process of determining how best to evaluate the quality of recommendations while taking into account the values and goals of each of these stakeholder groups. Qualitative research methods, such as interviews, focus groups, surveys \cite{kuniavsky2003observing}, contextual inquiry \cite{raven1996using}, and co-design \cite{steen2011benefits} could all be beneficial in this process.


\subsection{Values and Goals}\label{sec:methodology:values}
Once the stakeholders have been identified, the next step involves looking at the values they would want to be part of the recommendation task. Stakeholders' values are at the core of the evaluation process since they drive the modeling of the overall optimization problem. They represent high-level and abstract objectives the stakeholders wish to be satisfied via the use of the recommendation platform \cite{jannach2016recommendations}.
For instance, if the stakeholder is a music consumer, a possible value is usefulness (of music experience). Conversely, for music providers, a value could be monetary reward or (societal) well-being. It is worth noticing that values may also overlap or partially compete with each other. 

The elicitation of values is a fundamental (yet sometimes neglected) step, as it allows the actors involved in designing the system to formulate the goals of each stakeholder involved in a multistakeholder scenario. Going back to the music consumer and provider in our hypothetical example, possible goals might be accuracy and diversity of the recommendation results for the consumer, sell as many items or services as possible, grow the number of users, sell elements over the whole catalog, protect underrepresented groups, or reduce the carbon footprint for the provider. Different from values, goals can be tailored to the specific recommendation domain. A provider may set its goal to grow the number of users listening to classical music, and a consumer may wish to have diverse song recommendation with respect to genre. Goals are more detailed and measurable objectives than values, and they drive the design and implementation of the system through specific evaluation metrics.  


\subsection{Evaluation Metrics}\label{sec:methodology:metrics}
Formal evaluation metrics provide a way to measure the extent to which the goals of various stakeholders are achieved, i.e., they are measurable proxies towards goals. For example, both consumers and providers are likely to be interested in recommendation accuracy, consumers may be further interested in item discoverability (diversity, novelty, coverage); providers are likely interested in increasing revenue and engagement; and, the third-party stakeholders (for instance, regulators) are likely to be interested in protection-related metrics for consumers and providers (representation, fairness, etc.).

Multiple metrics can measure the success of the same goal, depending on the point of view or the aspect we want to highlight. For example, there are different metrics to measure accuracy, e.g., Normalized Discounted Cumulative Gain (NDCG), Mean Reciprocal Rank (MRR), or Recall; we may measure the overall number of items sold in a specific period or a specific geographical area, the items from the long-tail and the short-head, etc. Depending on the goal, we may have metrics subset of targeting the overall population of users and stakeholders available in the system.

Some of the specific metrics will naturally come from the prior research literature in recommender systems---the reader may refer to \cite{gunawardana2022evaluating} for discussions of some best practices and key metrics in recommender systems evaluation. However, there are clearly opportunities for further metric design, especially for provider-oriented and third-party-oriented stakeholders (i.e., stakeholders that have been under-explored in recommender systems research). Some metrics of these types have been explored in fairness-aware recommendation research \cite{ekstrand2022fairness}.  All the metrics must be validated by the target stakeholders (a relevant subset of the overall population is sufficient) to check if they are actually representative of their goals and if they are able to differentiate between high and low utility results. Stakeholders involved in validating the metrics are asked to assess the meaningfulness of the computed results, compared to their goals. A further result of this validation process by the stakeholder can be that of identifying a priority among the metrics. Especially in this phase, a desirable characteristic of a metric is its interpretability and its propensity towards the generation of a human-readable explanation.



As the result of this step, a list of important evaluation metrics $(m_1, \ldots, m_n)$ is enumerated, which represents the set of important considerations across multiple stakeholders that need to be taken into account as part of the multistakeholder recommender system evaluation. 

\subsection{Strategies for Aggregate Evaluation}\label{sec:aggregation}
Identifying the list of important evaluation metrics $(m_1, \ldots, m_n)$, as discussed above, provides the ability to evaluate (i.e., to score) a given recommender system $R$ in a multidimensional manner. More formally, $\mathbf{S}(R) = (s_1, \ldots, s_n)$, where $s_i$ is the performance of $R$ with respect to measure $m_i$, i.e., $s_i = m_i(R)$. For some applications, this may be sufficient, as it provides a multidimensional view of the system's performance.

In other cases, it may be desirable to formulate a single aggregate measure that captures the overall (i.e., multistakeholder, multiobjective) performance of the system \cite{DBLP:journals/ijon/ZhengW22}. In particular, given two candidate recommender systems $R_A$ and $R_B$, where each of which can be evaluated according to the stated list of metrics, $\mathbf{S}(R_A)$ and $\mathbf{S}(R_B)$, how to design a multistakeholder / multiobjective evaluation mechanism $\prec_M$ that allows to determine whether system $R_B$ has superior overall performance to system $R_A$, i.e., $\mathbf{S}(R_A) \prec_M \mathbf{S}(R_B)$?  Designing multistakeholder / multiobjective evaluation mechanisms constitutes an important and non-trivial task, because the recommender systems (and ML-based predictive models, more generally) often exhibit significant tradeoffs when it comes to different performance metrics.  To mention a few examples, prior literature has shown that there are tradeoffs between accuracy and item ``discoverability'' measures (i.e., diversity, serendipity, novelty) \cite{AdomaviciusK12,CastellsHV22,10.1145/1125451.1125659}, between accuracy and fairness \cite{Corbett-DaviesP17,10.1287/mnsc.2021.4065, mehrotra2018towards}, between different fairness measures themselves \cite{Binns20,KleinbergMR17}, and that business-oriented metrics may not be always aligned with the accuracy-oriented metrics \cite{Gomez-UribeH16,jannach2017vams,10.1145/3370082}.  

From a high-level perspective, strategies for developing aggregate multistakeholder / multiobjective evaluation mechanisms $\prec_M$ can be classified into two general paradigms:
\begin{itemize}
\item {\bf Direct/user-specified multistakeholder evaluation mechanisms}, which typically rely on domain expertise to specify {\em explicitly} the relative importance and/or tradeoffs between individual performance metrics, which then can be used to formally define the overall multistakeholder evaluation mechanism $\prec_M$. Example strategies include:
\begin{itemize}
    \item Weighted (typically linear) aggregation of individual metrics \cite{DBLP:conf/ppsn/BrankeDDO04,1084969} into a single numeric score (as an overall performance), which then allows for a more straightforward comparison of candidate systems. 
    \item Reduction of metric dimensionality by converting some of the individual metrics into constraints \cite{4308298}. Constraints can be of various types, e.g., hard vs.\ soft constraints. Hard constraints may indicate the system performance requirements that must be satisfied, which then can be used to filter out candidate systems with inadequate performance.  Soft constraints may indicate the relative importance (prioritization) of some metrics, which then can be used to rank the candidate systems accordingly. 
\end{itemize}
\item {\bf Learning-based multistakeholder evaluation mechanisms}, which typically {\em infer} the overall multistakeholder evaluation mechanism $\prec_M$ from certain available information.  Example strategies include:
\begin{itemize}
    \item Determining the Pareto frontier of the multidimensional performance vectors of different candidate systems, and measuring the overall performance of a given system as its distance from the Pareto frontier \cite{DBLP:conf/emo/Fleischer03}.  One key consideration is specifying an appropriate distance metric for multidimensional performance vectors $(s_1, \ldots, s_n)$. 
    \item Learning $\prec_M$ from ``ground truth'' examples. This could be achieved by providing multiple examples of multidimensional performance vectors $\mathbf{S}(R_i)$ to domain experts, asking them to provide the ``ground-truth'' judgments regarding the overall performance, and then using machine learning techniques to learn the relationships between the individual metrics and overall performance.  For instance, the domain experts could rank pairs of performance vectors at a time, $\mathbf{S}(R_A)$ and $\mathbf{S}(R_B)$, and provide a ground-truth judgment of whether $\mathbf{S}(R_A) \prec_M \mathbf{S}(R_B)$ or $\mathbf{S}(R_B) \prec_M \mathbf{S}(R_A)$ (or neither, $\mathbf{S}(R_A) \approx_M \mathbf{S}(R_B)$). Learning-to-rank techniques can then be used to build a model for estimating $\prec_M$ from such training data.
\end{itemize}
\end{itemize}

More generally, development of multistakeholder / multiobjective evaluation mechanisms $\prec_M$ for recommender systems has connections to several rich research literatures, including multiobjective/multi-criteria optimization \cite{Ehrgott2005,Miettinen1998}, multi-criteria decision-making \cite{Triantaphyllou2000} (including its various methodologies, such as data envelopment analysis \cite{Charnes1995DEA}, conjoint analysis \cite{Green1990ConjointAnalysis}, multi-attribute utility theory \cite{Keeney1993}), machine learning \cite{DBLP:journals/corr/abs-2010-04104}, and possibly others, which provide promising directions for further research. 

Finally, once the multistakeholder evaluation mechanism is chosen/finalized, incorporating it into the overall recommender-system-based solution can be done in two general  ways. (1) A straightforward {\em post-processing} approach would be to generate a number of different candidate recommender systems instances (e.g., using different recommendation algorithm configurations) and to use the multistakeholder evaluation mechanism to select the best-performing system among them. (2) In contrast, the {\em algorithmic tuning} approach would ``push'' the multistakeholder evaluation mechanism into the recommendation process, where a recommendation algorithm uses multistakeholder considerations more directly as part of the recommendation calculation/optimization process. 

\begin{figure}[tbh]
    \centering
    \includegraphics[width=0.75\linewidth]{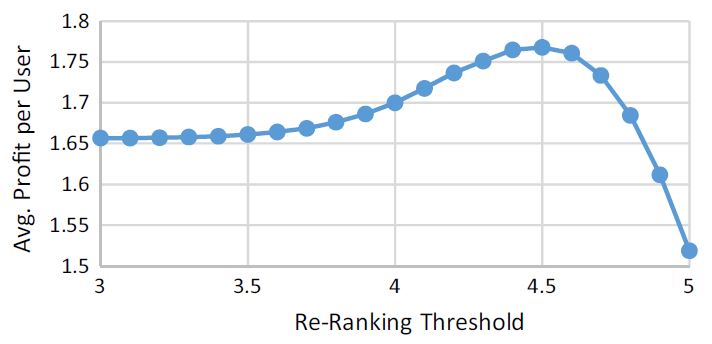}
    \caption{Recommendation relevance vs. profit (adapted from \cite{abdollahpouri2020multistakeholder,jannach2017vams})}
    \label{fig:profit_aware}
\end{figure}

As an illustrative example, Figure~\ref{fig:profit_aware} (adapted from \cite{abdollahpouri2020multistakeholder,jannach2017vams}) shows the performance of a simulated recommender system that takes into account the following multistakeholder considerations: item profitability (a key consideration for the recommendation provider) and recommendation relevance (a key consideration for the recommendation consumer). There is an inherent nuanced tradeoff between the two considerations. If the system is allowed to recommend only from the small number of the most relevant items (say, any items with predicted ratings above 4.9) for each user, the item relevance will be high, but choosing the most profitable items from only a small number of candidates will not improve the profitability metric. On the other extreme, if the system is allowed to recommend a much broader set of items (say, any items with predicted ratings above 3.0), the system will indeed be able to find more profitable items for recommendation, but the user may end up not choosing some of them due to their potentially low relevance (and may also be turned off by low-quality recommendations in the long run), thus, resulting in a lower actual profitability.  The system uses the algorithmic tuning approach to determine the relevance constraint/threshold (i.e., 4.5) that maximizes the profit metric, taking into account both the individual item profitability and the fact that the users will reject inaccurate recommendations. 

Additional considerations impacting the process of overall multistakeholder evaluation include:
\begin{itemize}
    \item \textbf{Stakeholder involvement.} Most of the aforementioned approaches will likely require the involvement of key stakeholders and domain experts, e.g., for determining tradeoffs between individual metrics (leading to decisions regarding relative importance weights for individual metrics or for determining which metrics should be converted to constraints), for obtaining ground-truth judgments about the overall system performance, etc. Therefore, one promising research direction is in the development of {\em participatory} frameworks \cite{WeBuildAI2019} that can enable and facilitate stakeholder groups to build algorithmic governance policies for computational decision-making and decision-support systems. 

    
    \item \textbf{Average vs. subgroup vs. individual performance.} It is imperative to establish the perspective for evaluation: Do we evaluate systems in terms of their average performance, or should the distribution of individual performance also be taken into account \cite{DBLP:conf/cikm/PaparellaAN0N23}?  For example, does higher average performance also come with much higher individual performance variance (i.e., much worse individual performance for some users/items/etc.), and, if so, what are the right trade-offs? More generally, evaluation at multiple granularities (various subgroup levels) may be of interest.
\end{itemize}



\subsection{Practical guidelines}\label{sec:methodology:guidelines}
Throughout this section, we have described in detail the best meta-practices for conducting successful multistakeholder evaluation, divided over different stages. We summarize these stages in the list below:

\begin{itemize}

    \item \textbf{Identification of stakeholders.} The inclusion of all relevant stakeholders is essential to the success and representativeness of evaluating a recommender system. Starting with the system stakeholders, consumers and producers, it is important to identify and involve all relevant downstream, upstream and third-party stakeholders. Researchers should consider a range of qualitative research methods and surveys, as well as literature reviews and low-fidelity design processes in doing so.

    \item \textbf{Articulation of values and goals.} The goals of the recommender system and expectations for what makes a recommendation good may vary considerably by stakeholder and context. Each stakeholder has different goals and expectations for the recommender system, and these are directly or indirectly tied to the values that matter to them. Qualitative research methods are particularly useful for identifying these values and goals. 
    It is important to keep domain differences in mind when identifying the values and goals. In some domains, certain stakeholders might be affected more seriously by the selection of recommendations provided by the system. It is therefore recommended to undertake a risk analysis at this stage to properly understand the extent of these risks.
    

    \item \textbf{Selection of evaluation metrics.} Once values and goals have been mapped, they have to be represented by measurable entities, e.g., by selecting existing metrics from prior research that are relevant for a given application context or by designing new ones. Part of this process includes determining which metrics to prioritize in the evaluation of the system, as typically multiple metrics can measure the success of the same goal depending on which perspective to highlight. These metrics can vary considerably in terms of being more qualitative or quantitative in nature, or in terms of representing more short-term or long-term interests. All selected metrics must be validated by the target stakeholders to check whether they are representative of their goals.

    \item \textbf{Strategy selection for overall multistakeholder evaluation.} Choosing a multistakeholder approach to recommender systems evaluation may also entail developing a strategy for creating an overall summative evaluation that integrates over the stakeholder perspectives. In other words, evaluating overall performance in a multistakeholder system means that we typically have to deal with a multitude of evaluation metrics, which could also substantially differ between the stakeholder groups. For example, if we know the system performs well for consumer-side metrics, how does this version of the system now work for provider-side metrics? Different strategies for evaluating the overall multistakeholder performance can be employed to find a suitable solution as discussed in Section~\ref{sec:aggregation}.
    
\end{itemize}

Also, from the practical perspective, the multistakeholder evaluation methodology---the identification of key stakeholders and their values/goals, the choice of most appropriate individual metrics, the development of specific multistakeholder / multiobjective evaluation mechanisms, and the use of these mechanisms to guide system design and improvement---can be viewed as an iterative process, where researchers and system designers should be aware of all the key steps and can return to iteratively refine any of them.

In reporting on multistakeholder recommendation research, we encourage researchers to include in their discussion the details of stakeholder identification and consultation, the derivation of values and goals, and the justification of metrics in terms of that work. \citet{selbst2019fairness} make the point that formalizations developed in addressing one problem do not necessarily transfer to other contexts. The authors were writing in the context of machine learning fairness, but multistakeholder recommendation is also highly context-specific and similar principles apply.

\section{Example Applications and Metrics}
\label{sec:multis:examples}

Deriving an evaluation metric requires working from a construct, an abstract quality of the recommendation process that we would like to understand, to a concrete proxy of that construct that can be measured and designing a methodology to measure it. The application-specificity of multistakeholder evaluation makes it difficult to provide such analysis in a general way. With that in mind, we present several specific examples, which illustrate the design process for a variety of metrics for different stakeholder in different applications. 

In these examples, we are not seeking to develop a full set of metrics for given application. As we note above, executing all of the stages of the evaluation design process can be a lengthy endeavor requiring detailed stakeholder consultation. For our hypothetical examples, we consider one thread of that much more involved process, focusing on a particular stakeholder, a specific value that they may have, and a particular metric that could be developed to evaluate it. It is worth reiterating that with these examples, we neither aim to provide a complete set of metrics that one might wish to implement in each of these settings nor to highlight the most important metrics. Rather, we seek to illustrate the type of analysis needed to derive such metrics. Moreover, we expect the process of metric selection and development to be iterative rather than linear; this process may take multiple rounds of consultation and implementation to derive a metric (or set of metrics) that captures a particular stakeholder's perspective.

The three areas chosen are music streaming, educational resources, and job recommendations. These examples were chosen to highlight different stakeholder perspectives. In music streaming, we focus on musical artists as an example of the provider role. In the recommendation of educational materials, we have a domain where the value of a recommender is more than just consumer taste and yet personalization is still important; here we focus on the student, a consumer role. Job recommendation is, in many countries, subject to regulation intended to ensure non-discrimination in hiring and is therefore a good place to explore evaluation from the perspective of third-party stakeholders.

\subsection{Music Streaming}

The first example we consider is streaming music recommendation with the key stakeholders introduced in Figure~\ref{fig:ms-example-music}, and also included in  Table~\ref{tab:stakeholders-music}.

In this case, we focus on the providers: the musical artists. There are a variety of values that such individuals might have concerning a distribution platform like a streaming service. We concentrate on the construct of \textbf{audience}: an artist will often seek to build a community of individuals who appreciate their particular musical style and contribution (connection, community and social bonding) and might, for example, come to a concert or purchase merchandise (monetary reward) in addition to listening through the streaming service. 

A given musical artist might seek to understand to what extent is the recommender system helping them build an audience (use value). One can imagine the system failing in various ways. It might recommend their music to listeners interested in something else, and so the recommendations are not acted upon. Or it might recommend the artist's music only to listeners who are already fans: helping cement the audience, but not necessarily building it over time. True audience building might only be evident over a long period of time (repeating habitual listening, ticket and merchandise purchases, etc.) so it will probably be necessary to create a short-term proxy for the audience-building potential of a recommender system (growth and market development). 

As this is a hypothetical example, our metric is necessarily speculative, but again the aim is to illustrate a process for developing such metrics, not to solve a given evaluation problem. First, we have the problem of measuring an audience from the data available within the streaming service. Let $r$ be the musical artist and let listen count $k_u = \ell(r, u, t)$ be the number of times that user $u$ listens to a track by $r$ over some standard time window $t$, perhaps one month. The audience $A_r$ can then be defined as the set of individuals for whom this count is greater than some threshold $\epsilon$: $k_u > \epsilon$. 

As noted above, measuring audience development can have a long time scale, so a short term proxy for this quality could be to measure to what extent an artist's music is being recommended to receptive users. There are multiple ways to determine if a user is receptive\footnote{For example, did the user listen to a second song by the artist, add their songs to a playlist, etc.?}, but for the sake of example, let us assume that we can measure the number $n$ of non-audience listeners (that is, $u \notin A_r$) who were recommended a song by $r$ and then listened to the entire song. Given that different musicians have very different numbers of fans, it might make sense to normalize by the size of the artist's existing audience $A_r$: $m_r = n/|A_r|$. 

As a metric shared with individual providers, a low score on $m_r$ might raise concerns for the artist relative to the recommender system. It would mean that few new listeners are being introduced to their music. For a superstar, this might not be an issue: many people know their music already, but for an emerging artist, it could indicate that the recommender is not as helpful as it might be. A higher  $m_r$ score does not necessarily mean that their audience is growing, but it does mean that the recommender system is introducing their music to potential new fans. From the system stakeholder point of view, this score could also be aggregated across all providers to understand audience building across the platform's stable of artists. Its distribution might also be relevant in terms of fairness: are some types of artists better able to build audiences on the platform than others? 

\subsection{Education}

In the context of educational recommender systems, our example focuses on a course content recommender system for secondary school students, possibly integrated within a learning management system (LMS) where the system could track the progress of each student and generate recommendations about what to study next. 
We illustrate the relationship between value-driven goals and potential measures of each stakeholder, and show how the evaluation perspective changes according to the goal in focus.

In this scenario, teachers provide the content to the recommender system platform both by selecting relevant external content (e.g., educational videos, reference books and articles) and content generated by themselves. Therefore, we define the external content generators as upstream stakeholders and teachers as provider stakeholders.

The recommender system platform generates course content recommendations for students who are consumer stakeholders and direct users of the system. Parents of the students have an indirect relationship with the generated content (e.g., in the context of recommendation of educational materials for secondary school students, parents might be interested in checking the type of material their children are using) and they are defined as downstream stakeholders. Both upstream and downstream stakeholders have an indirect relationship to the RS platform, which may be relevant to identify and evaluate the value driven goals in a greater picture.

\begin{figure}[tbh]
    \centering
    \includegraphics[width=1\linewidth]{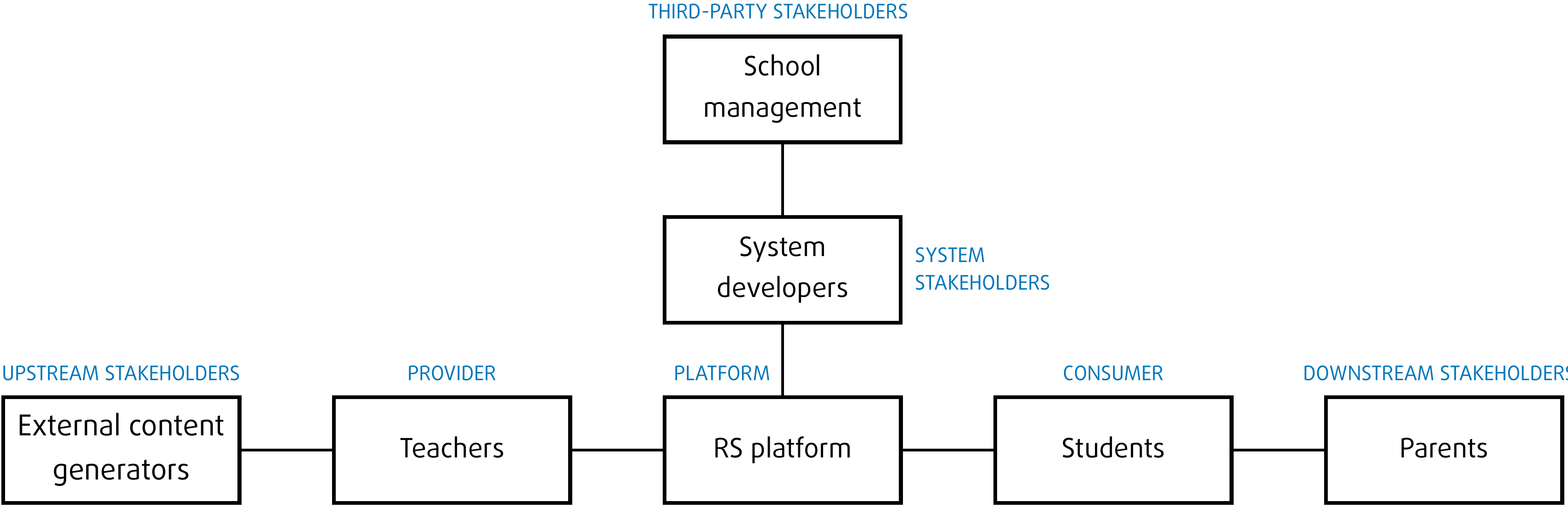}
    \caption{Stakeholder configuration for the education example}
    \label{fig:multistakeholder_relations}
\end{figure}

The system stakeholders are responsible for the seamless operation of the recommender system, and they are obliged to ensure that the recommender system platform follows the laws and regulations stated by the school management who is among the {\em third-party} stakeholders (e.g., the recommended content should be within the corresponding curriculum for each student). Figure~\ref{fig:multistakeholder_relations} illustrates the multistakeholder relations, goals and potential measures in this example scenario.

Based on this example scenario, one point of evaluation of the recommender system platform could be done from the perspective of one of the goals of the consumer stakeholder. More specifically, we could evaluate the recommender system platform from the students' perspective of passing a course, answering the question 
``How likely is it that a student passes a course when she follows the recommendations from the platform?'' (usefulness and enjoyment, as well as personal growth). 
Although defined from the recommendation consumer's perspective, other stakeholders may benefit the same evaluation. For example, the teacher could use the same measure to understand if the resources she provided to the platform are sufficient in type and quality (usefulness and enjoyment), and the system developers might get an understanding of the relevancy of the recommendations generated by the system beyond click-through rate (use value).

Since the goal of the student is to pass the course at the end of the semester, in this example, we need to evaluate our system at the end of each semester. We assume that the student $S_i$ receives $n$ recommendations every time she uses the system. $S_i$ may choose to accept a recommendation or do another activity on the platform. Therefore, we can measure the number of accepted recommendations by student $S_i$ throughout the semester being $n_i$. The acceptance of recommendations can be measured in different ways, but for the sake of this example, if the student clicks on any of the recommendations on the list, we assume that the recommendation has been accepted. $k_i$ being the total interaction count of $S_i$ with the system, we can calculate the proportion of the accepted recommendations to the number of whole interactions as $p_i$=$k_i$/$n_i$. Finally, at the end of the semester, we calculate the correlation between the student’s final grade in the course and $p_i$.
For the sake of this example, we skip the importance of the order of the recommendations, but an evaluation metric such as NDCG could easily be employed for this purpose. Further, the final metric that correlates the acceptance of recommendations with the student’s final score, could be calculated based on the order of the recommendations, answering the question: Does accepting higher-ranked recommendations from the list correlate with higher student scores?\footnote{One might argue for a different indicator of educational value---perhaps the student's understanding is enhanced in ways less directly measurable---but this equation of final grade with educational value is common in the literature.}


We should note that the goals of each student may differ or we might be able to identify clusters of students who share the same goals. Therefore, the evaluation methodology could be adjusted according to not only different types of stakeholders but also the differences within one type of stakeholder. This concept of granularity has been discussed in Section \ref{ms-methodology}. 
Similarly, different stakeholders may have different temporal requirements based on their goals. For example, the students may have a goal for the whole semester (e.g. passing the course), whereas the teachers may have goals that need to be evaluated in a shorter term (e.g. understanding if the recommender system platform is helpful for the students to understand the weekly topics).

\begin{table}[tbh]
\resizebox{1\textwidth}{!}{%
\begin{tabular}{|l||l|l|l|l|l|l|}
\hline
 &
  \textbf{Upstream} &
  \textbf{Provider} &
  \textbf{System} &
  \textbf{Third party} &
  \textbf{Consumer} &
  \textbf{Downstream} \\ \hline
\textbf{Stakeholder} &
  \begin{tabular}[c]{@{}l@{}}External content \\ generators\end{tabular} &
  Teachers &
  RS platform &
  \begin{tabular}[c]{@{}l@{}}School\\ management\end{tabular} &
  Students &
  Parents \\ \hline
\textbf{Goals} &
  \begin{tabular}[c]{@{}l@{}}Economic gain, \\ reputation, social \\ benefit\end{tabular} &
  \begin{tabular}[c]{@{}l@{}}Educating younger \\ generation, social \\ benefit\end{tabular} &
  Economic gain &
  Social benefit &
  \begin{tabular}[c]{@{}l@{}}Passing the course, \\ learning\end{tabular} &
  \begin{tabular}[c]{@{}l@{}}Educating their \\ children\end{tabular} \\ \hline
\textbf{Measures} &
  \begin{tabular}[c]{@{}l@{}}Exposure, generating\\ high-quality content\end{tabular} &
  \begin{tabular}[c]{@{}l@{}}Students learning \\ well, generating \\ high-quality content\end{tabular} &
  \begin{tabular}[c]{@{}l@{}}Ensuring that the \\ RS works properly, \\ ensuring that the \\ requirements from \\ other stakeholders \\ are satisfied\end{tabular} &
  \begin{tabular}[c]{@{}l@{}}Ensure that laws \\ and regulations \\ are being followed\end{tabular} &
  \begin{tabular}[c]{@{}l@{}}Getting good grades, \\ learning the topics \\ well\end{tabular} &
  \begin{tabular}[c]{@{}l@{}}Reviewing the course \\ material, giving advice \\ to their children\end{tabular} \\ \hline
\end{tabular}
}
\caption{Sample stakeholder goals and measures for the education example}
\label{tab:stakeholders-ed}
\end{table}

\subsection{Human Resources}

\begin{figure}[tbh]
    \centering
    \includegraphics[width=1\linewidth]{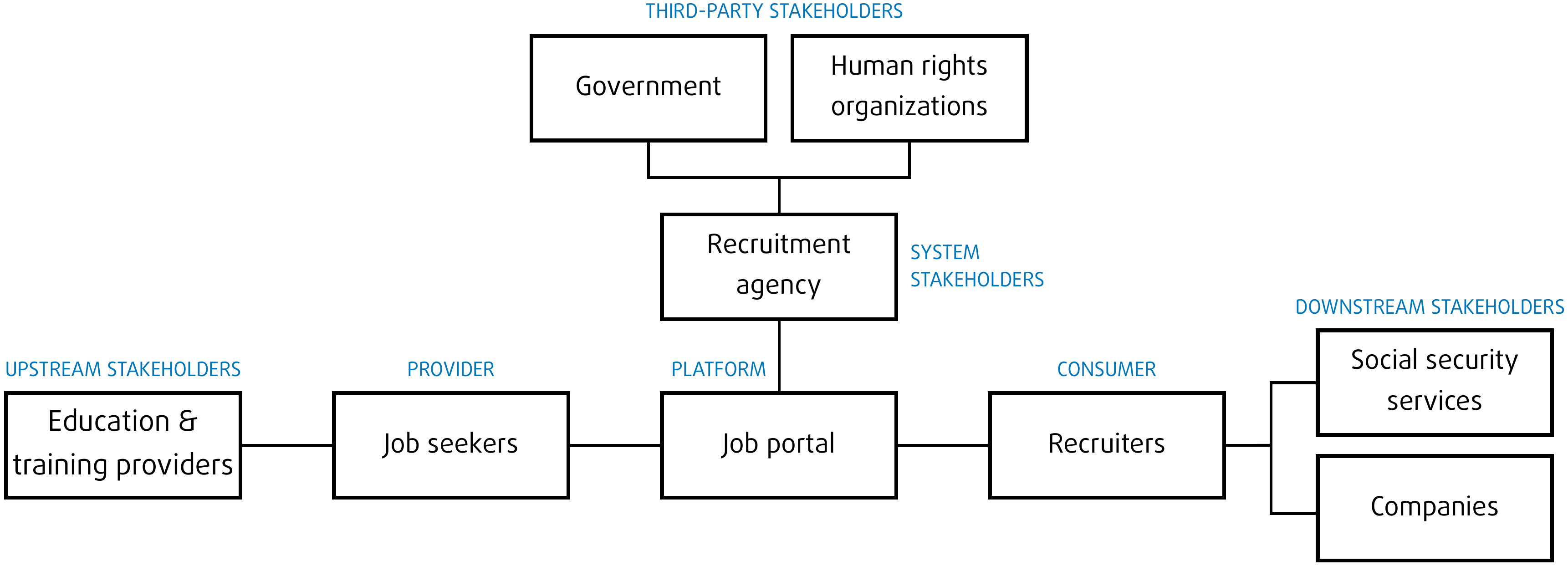}
    \caption{Stakeholder configuration for the human resources example}
    \label{fig:ms-example-jobs}
\end{figure}

The final example we consider is \textbf{candidate recommendation}: recommending suitable candidates for an open job position, also known as talent search. 
Recruiters often play an important intermediary role in this process by assessing candidates' qualifications in relation to the job~\cite{Breaugh:2008}. The candidate identification and assessment process places a great manual burden on recruiters~\cite{Montuschi:2014} and they would benefit from a system that recommends relevant candidates to supplement their own manual searches.  
Figure \ref{fig:ms-example-jobs} illustrates the different stakeholders involved in this recommendation scenario and is supplemented by Table~\ref{tab:stakeholders-hr}, which displays example goals and measures for each of the stakeholder categories.

\begin{table}[tbh]
\resizebox{1\textwidth}{!}{%
\begin{tabular}{|l||l|l|l|l|l|l|}
\hline
 &
  \textbf{Upstream} &
  \textbf{Provider} &
  \textbf{System} &
  \textbf{Third party} &
  \textbf{Consumer} &
  \textbf{Downstream} \\ \hline
\textbf{Stakeholder} &
  \begin{tabular}[c]{@{}l@{}}Education \& \\ training providers\end{tabular} &
  Job seekers &
  Job portal &
  Government &
  Recruiters &
  Companies \\ \hline
\textbf{Goals} &
  \begin{tabular}[c]{@{}l@{}}Personal develop-\\ ment, monetary \\ reward\end{tabular} &
  \begin{tabular}[c]{@{}l@{}}Personal development, \\ well-being, monetary\\  reward, social bonding\end{tabular} &
  \begin{tabular}[c]{@{}l@{}}Monetary reward, \\ customer satisfaction, \\ customer loyalty\end{tabular} &
  \begin{tabular}[c]{@{}l@{}}Employment, social \\ cohesion, economic \\ development, quality \\ of life\end{tabular} &
  \begin{tabular}[c]{@{}l@{}}Recognition \& \\ acknowledgment, \\ personal autonomy, \\ well-being, social \\ bonding\end{tabular} &
  \begin{tabular}[c]{@{}l@{}}Monetary reward,\\ market develop-\\ ment, employee \\ well-being\end{tabular} \\ \hline
\textbf{Measures} &
  Grading scale &
  \begin{tabular}[c]{@{}l@{}}Salary increase, \\ working hours\end{tabular} &
  \begin{tabular}[c]{@{}l@{}}Response rate, \\ \% hired, time \\ spent per job, time \\ spent per candidate\end{tabular} &
  \begin{tabular}[c]{@{}l@{}}Unemployment rate, \\ GDP growth, \\ happiness index\end{tabular} &
  \begin{tabular}[c]{@{}l@{}}No. of queries \\ issued, time spent \\ per candidate, \\ time spent per job, \\ no. of candidates \\ contacted\end{tabular} &
  \begin{tabular}[c]{@{}l@{}}Time until position \\ is filled\end{tabular} \\ \hline
\end{tabular}
}
\caption{Sample stakeholder goals and measures for the human resources example}
\label{tab:stakeholders-hr}
\end{table}

This recommendation scenario starts with job seekers by signaling they are open to finding a new job by uploading their CV to the job portal's CV database, making them the item {provider stakeholder. In this scenario, the recruiter is the party receiving the recommendations, making them the consumer stakeholder. The system stakeholder is responsible for creating and operating the candidate recommender system on the job portal, which suggests a slate of relevant candidates to the recruiters. Their values are not necessarily the same as those of the customers and providers. Here we assume that the recruitment agency is the system stakeholder and that they are seeking making their recruiters more efficient through an effective recommender system that allows recruiters to complete job / candidate matches. 

Despite paying for the recruitment service, the company with the open job position is not a customer from a multistakeholder evaluation point of view. In this scenario, they instead play the role of downstream stakeholder, as they are impacted by the choices of the recruiters make when assessing, shortlisting and contacting the recommended candidates. 

Upstream stakeholders are those potentially impacted by the recommender system, but not direct contributors of items. In the candidate recommendation scenario, education and training providers could function as an upstream stakeholder. These education providers do not have a direct stake in the candidate recommender system but could be interested in learning which skills and competences are most important for a successful matching process, allowing them to update their programs and courses.

Government institutions are an example of third-party stakeholders: they do not have any direct interaction with the job portal, but they have an interest in or are impacted by its operation. A successful candidate recommender system could result in more successful matches between job seekers and companies, affecting important government values such as societal benefit, growth and market development, and well-being. 

Government institutions can also have a more direct impact on and interest in the job portal's operation through legislation that ensures non-discrimination in hiring practices, something shared by human rights organizations. Such regulatory practice may impose legally binding requirements on the system stakeholders, affecting the evaluation of the recommended slates of candidates in terms of fairness and protecting underrepresented groups. Job recommendation is therefore a good example to explore evaluation from the perspective of third-party stakeholders.

More specifically, we could evaluate the recommender system platform from the governmental perspective of fairness, answering the question ``Given a set of candidates qualified for a job, do the job seekers in both protected and unprotected groups have an equal probability of being contacted?'' This question matches the notion of group fairness (or statistical parity), one of the wide variety of fairness metrics \cite{garg2020fairness}. In our scenario, group fairness is defined as both protected and unprotected groups having an equal probability of being suggested to the recruiter by the recommender system, given they all meet the qualifications set out in the original job posting. Protected groups are defined in terms of sensitive attributes, such as gender, age, ethnicity, and sexual orientation. For example, if a legislative body wanted to ensure gender fairness, an evaluation metric based on group fairness would check whether the difference between the probability of being contacted from the protected group $P(\mathit{contacted} | \mathit{qualified} \land G \neq \mathit{male})$ is equal to the probability of being contacted in the unprotected group $P(\mathit{contacted} | \mathit{qualified} \land G = \mathit{male})$ is close to zero.\footnote{Note that assessing whether a given candidate matches the job qualification and to what degree may be complex task in itself.} In an actual multistakeholder evaluation, it would be essential to involve the other stakeholders in determining what fairness means for them, which sensitive attributes are relevant, and how to map this to the most relevant fairness metrics.

\section{Supporting Stakeholder Insights and Control}\label{sec:challenges}

A holistic approach to understanding and supporting the perspectives and needs of multiple stakeholders involves a range of challenges that go beyond what we have discussed so far. In the previous sections, we have described general properties of multistakeholder recommendation, and methodological approaches to developing relevant metrics, and outlined three hypothetical examples of metric development targeted to different classes of stakeholders. While several of these activities will require close cooperation and coordination with the different stakeholders, they do not yet address the question of how stakeholders can be empowered to gain insights relevant to them during system operation or even control some of the system's functionalities. While these questions have increasingly been investigated in recent years for consumers as the main stakeholder class, very little research has yet addressed how other stakeholder classes might be supported in accessing and understanding information about the recommender system that serves their specific needs. Furthermore, the question arises whether, and to what extent stakeholders should be enabled to exert influence on the recommendation process itself. In this section, we discuss some salient challenges: Transparency \& Explainability,  Strategic / Adversarial Considerations, Interfaces and Interactive Control, and Governance.  

\subsection{Transparency and Explainability}
Developing multistakeholder metrics and evaluation processes raises the question of to whom such metrics might be reported and made available. Recommender systems evaluation is typically a purely internal matter of engineers or system operators understanding how the recommender operates and seeking to improve it.

However, the types of evaluations that we discuss are different in that they may be of interest to parties who normally have no access to the workings of the recommender system. For example, the musical artists in our streaming example typically have very little insight into how the recommender system treats their content. While some major commercial music streaming platforms provide quite detailed analytics data for artists (see, e.g. Spotify \footnote{\url{https://artists.spotify.com/analytics}}), the specific influence of the recommender system with respect to audience building or fairness is typically not exposed to the artists.
However, we noted earlier that a musical artist might seek to understand to what extent the recommender system is helping them build an audience. Such  metrics could be shared with artists as a form of explanation to help them understand, for example, how frequently their music is recommended in comparison to other artists (possibly with a similar genre and style) .

Explanations in a multistakeholder context also bring challenges different from explanations targeted only toward consumers. Firstly, different stakeholder groups have different explanatory needs that need to be identified. In the aforementioned example, the artist, their listeners, or the music label have quite different explanatory needs. This leads to the question of presenting  different explanations to different stakeholders, or even personalizing them. For example, should we explain item recommendations to individual users and audience building to artists separately? When these concerns are integrated in an overall system objective, an additional type of explanation might be one that explains how we resolve the exposure of artists relative to how we weigh user preferences?
In other words, if the requirement is to find and generate such a general explanation, then we need systems that can generate a meta-explanation on how tensions were resolved. A particular tension has already been observed where consumers (depending on the context of recommendation) may prefer less transparency in explanations if it gives better privacy \cite{najafian2024people,zilbershtein2024bridging}. Members of groups preferred not to disclose sensitive information to other group members, and consumers of advertisements did not want certain information to be used in explanations (or as a basis of recommendations!).
We do not attempt to answer the question of how to generate these kinds of explanations (as this is out of scope), but note that provider-side transparency, let alone generalized explanations, are very little studied in the context of multistakeholder recommendation.

\subsection{Strategic / Adversarial Considerations}

One likely reason that multistakeholder transparency has been little pursued is the concern that such a facility might be used to enable undesirable adversarial behavior. A web search for the term ``YouTube algorithm'' yields thousands of hits from search engine optimization (SEO) firms and others advising creators about how to bend the algorithm to their will. Additional information given to providers may enhance their ability to manipulate the algorithm in ways that are not necessarily beneficial to recommendation consumers or the platform. An open research question is how to offer provider-side disclosure in a way that limits adversarial opportunities. Where it is not possible to provide such disclosure, platforms may still seek to measure provider-side properties of their systems, but only for internal consumption. 



\subsection{Interfaces and Interactive Control}

When translating transparency and explanation goals into concrete system designs, the question arises of how different classes of stakeholders can interact with the recommender system. There is a great deal of study of consumer-side recommendation interfaces, and a wide variety of interface designs for presenting recommendations, and for generating and interacting with recommendations. Recommender systems interfaces for other stakeholders do exist but are rarely the subject of published research. For example, YouTube provides a set of tools within their YouTube Studio application\footnote{\url{studio.youtube.com}} to enable video creators to see some information about the viewership of their videos, but there are no detailed analytics about how the recommender system is handling their content or ways to interact with the recommender system itself. Various suppliers of recommender system software or services also provide functionality for monitoring the recommendation  process, mostly in the form of GUI-style dashboards. Depending on the platform, these dashboards present different types of analytic data, such as the number of personalized and non-personalized recommendations, the number of user interactions, or the temporal development of such data. However, the main target of monitoring dashboards are platform providers. Upstream or downstream stakeholders are generally not accommodated. Moreover, no specific evaluation approaches have become available so far for non-consumer stakeholders.  

Today, consumers are used to static one-shot recommendations that cannot be influenced by them. In contrast, interactive and conversational systems are emerging as methods that can provide users with different levels of control over the recommendation process, turning the one-shot process into a multi-turn dialog (\cite{DBLP:journals/csur/JannachMCC21}. Examples of interactive methods include, for example, selecting or weighting the input data used by the algorithm,  changing how strongly different item features influence the recommendation outcome, or interactively critiquing sample items. As we have noted, system and provider objectives may be quite different from those of consumers. We may expect that interactive recommendation techniques may strongly privilege the consumer side of the recommendation interaction; the impacts on other stakeholders of these technologies is still a matter for future research. Although methods for interacting with recommenders have mostly been GUI-based, the recent advances in language technologies have made text- or speech-based conversational approaches very promising. This holds both for controlling the recommendation process and for providing explanatory information \cite{hernandez-bocanegra_explaining_2023}. However, conversational techniques have thus far also only been directed at consumers, but applying their capabilities to other stakeholder groups would provide these with flexible means to query or control the system according to their specific needs. 

In parallel with lack of attention on interfaces and interaction for non-consumer stakeholders, specific evaluation metrics and methods are rare or missing, although the interactive process itself may strongly affect the satisfaction of some of the stakeholders' goals. Existing interaction-related metrics such as  the number of interactions to get the final recommendation \cite{DBLP:journals/isci/NoiaDJNP22} or the seamless perception of the interactive process \cite{DBLP:conf/intrs/ManzoorCJ23}, solely address the consumer side. There is also little development of user- or system-oriented metrics specifically for conversational recommendation techniques. As a result, interaction is an underexplored aspect of multistakeholder recommender systems. Except for a few recent qualitative studies \cite{choi2023creator,smith2024recommendme}, there is relatively little knowledge about techniques and user experiences with recommender system interfaces for stakeholders along the entire value chain.

\subsection{Governance}

Our aim with this article is to help researchers and system designers consider more holistic evaluations of recommender systems, taking multiple stakeholders into account, and examining the impact of the system across stakeholder groups. There is a separate question of governance: who, in the end, has a concrete and effective say in how a recommender system operates? Corporate structures often have a very concrete answer to this question, but as media scholar Nathan Schneider reminds us \cite{schneider2024governable}, other models of governance can be and have been applied to online systems. Multistakeholder governance of recommender systems is an interesting question for future research and development. 

\section{Conclusions}

Multistakeholder recommender systems account for the impacts and preferences of multiple groups of individuals, not just the end users receiving recommendations. This article focuses on the evaluation of such multistakeholder recommender systems. We provide examples from three domains (music, HR, and education). Additionally, we discuss how to move from theoretical principles to practical implementation. We bring attention to the different aspects involved---from the range of stakeholders involved (including but not limited to producers and consumers) to the values and specific goals of each relevant stakeholder. Next, we discuss evaluation methodologies and metric selection. Finally, we outline several open areas for future research, as seen from the lens of multistakeholder evaluation. 

The structure of the paper follows the steps necessary to evaluate a methodology for software application development through: defining relevant features; scoring the features; and providing guidelines for applying the evaluation framework \cite{10.1145/251880.251912}. Overall, with this work, we aimed to outline the components that cannot  be overlooked when undertaking a multistakeholder evaluation of recommender systems. As we have emphasized throughout the narrative, it cannot be denied that each use case is unique, influenced by the characteristics of the primary recommender system, the environment in which it is deployed, and the (direct and indirect) users it is designed to serve. Nevertheless, the components we discuss, along with the focus on the practical implementation of key areas—from identifying stakeholders and eliciting their views on recommender systems to defining and measuring how well the envisioned values for a recommender are upheld—should serve as guidance for researchers and practitioners. This understanding is essential for navigating the complex task of evaluating recommender systems that satisfy multiple stakeholders while considering the necessary trade-offs and compromises that need to be made along the way.

\section*{Acknowledgments}
The authors thank the Schloss Dagstuhl – Leibniz Center for Informatics for sponsoring and hosting the Evaluation Perspectives of Recommender Systems: Driving Research and Education seminar, and we also thank Christine Bauer, Alan Said and Evan Zangerle for organizing the seminar. \\
This publication is supported by the ROBUST project: Trustworthy AI-based Systems for Sustainable Growth with project number  KICH3.LTP.20.006, which is (partly) financed by the Dutch Research Council (NWO), RTL, DPG, and the Dutch Ministry of Economic Affairs and Climate Policy (EZK) under the program LTP KIC 2020-2024. \\
The financial support by the Austrian Federal Ministry of Labour and Economy, the National Foundation for Research, Technology and Development and the Christian Doppler Research Association for research by Julia Neidhardt is gratefully acknowledged.\\ 
The work of Robin Burke was supported by the National Science Foundation under award IIS-2107577. \\
The work by Toine Bogers was supported by the FairMatch project (Innovation Fund Denmark grant number 3195-00003B) and by the Pioneer Centre for AI (DNRF grant number P1).\\
The work by Dominik Kowald was supported by the Austrian FFG COMET – Competence Centers for Excellent Technologies Program and funded by BMK, BMAW, as well as the co-financing provinces Styria, Vienna and Tyrol.\\
\textit{All content represents the opinion of the authors, which is not necessarily shared or endorsed by their respective employers and/or sponsors.}